\RequirePackage{lineno}
\documentclass[final, 3p, times, twocolumn]{elsarticle}
\usepackage{float}
\usepackage{comment}
\usepackage{epsfig}
\usepackage{amssymb}
\usepackage{bm}
\usepackage{graphicx,color}
\usepackage{dcolumn}
\usepackage{longtable}
\usepackage{epstopdf}
\usepackage{csquotes}
\usepackage{amsmath}
\usepackage{physics}
\usepackage{textgreek}
\usepackage{placeins}
\usepackage{dblfloatfix}
\usepackage{rotating}
\usepackage{subcaption}
\usepackage{hyperref} 
\usepackage[table]{xcolor}

\begin{document}
\title{Development of a Modular Current-Mode NaI(Tl) Detector Array for Parity Odd (n,$\gamma$) Cross Section Measurements}

\author[1,2]{J.T. Mills \corref{cor1}}

\author[1]{J. G. Otero Munoz \corref{cor1}}

\author[1]{K. Dickerson}

\author[2]{I. Britt}

\author[3]{A. Couture}

\author[1]{J. Doskow}

\author[2]{J. Fry}

\author[7]{I. Ide}
\author[7]{M. Kitaguchi}
\author[6]{R. Kobayashi}

\author[1]{M. Luxnat}

\author[4]{A. Moseley}

\author[7]{R. Nakabe}

\author[5]{I. Novikov}

\author[6]{K. Oikawa}
\author[6,11]{T. Oku}
\author[7,6]{T. Okudaira}

\author[2]{A. Quintinar-Pe\~na}

\author[2]{A. Richburg}

\author[1]{S. Samiei}

\author[1]{D. Schaper}

\author[7]{H. M. Shimizu}

\author[2]{D. Slone}

\author[1]{W. M. Snow}

\author[6,8]{S. Takada}
\author[6,9]{S. Takahashi}
\author[6]{Y. Tsuchikawa}

\author[1]{G. Visser}

\author[3]{J. Winkelbauer}

\affiliation[1]{organization={Indiana University},
addressline={107 S Indiana Ave},
city={Bloomington},
postcode={47405},
state={IN},
country={USA}}

\affiliation[2]{organization={Eastern Kentucky University},
addressline={521 Lancaster Ave.},
city={Richmond},
postcode={40475},
state={KY},
country={USA}}

\affiliation[3]{organization={Los Alamos National Laboratory},
addressline={P.O. Box 1663},
city={Los Alamos},
postcode={87545},
state={NM},
country={USA}}

\affiliation[4]{organization={University of Kentucky},
addressline={506 Library Dr},
city={Lexington},
postcode={40508},
state={KY},
country={USA}}

\affiliation[5]{organization={Western Kentucky University},
addressline={1906 College Heights Blvd.},
city={Bowling Green},
postcode={42101-3576},
state={KY},
country={USA}}


\affiliation[6]{organization={Japan Atomic Energy Agency},
addressline={2-4 Shirakata, Tokai},
city={Ibaraki},
postcode={319-1195},
country={Japan}}
\affiliation[7]{organization={Nagoya University},
addressline={Furo-cho, Chikusa},
city={Nagoya},
postcode={464-8602},
country={Japan}}
\affiliation[8]{organization={Tohoku University},
addressline={2-1-1 Katahira, Aoba-ku},
city={Sendai},
postcode={980-8576},
country={Japan}}
\affiliation[9]{organization={The University of Tokyo},
addressline={5-1-5 Kashiwanoha, Kashiwa},
city={Chiba},
postcode={277-8581},
country={Japan}}
\affiliation[11]{organization={Ibaraki University},
addressline={2-1-1 Bunkyo, Mito},
city={Ibaraki},
postcode={310-8512},
country={Japan}}

\cortext[cor1]{Corresponding author. These authors contributed equally to this work.}

\date{\today}

\begin{abstract} 
The Neutron Optics Parity and Time-Reversal Violation Experiment (NOPTREX) Collaboration has developed a modular array of 24 NaI(Tl) detectors to measure parity and time-reversal symmetry violation in neutron-nucleus interactions. These detectors feature custom electronics that allow for operation in pulse or current mode. This paper describes the design, construction, characterization, and testing of the detectors in this array. We demonstrate the ability of the array to detect parity-odd asymmetries in neutron resonances by observing the known 0.7 eV parity-violating resonance in $^{139}$La in measurements at LANSCE.
\end{abstract}

\maketitle

\section{Introduction}




Parity violation (PV) in neutron-nucleus interactions provides a unique probe of the weak interaction between hadrons. In general, the parity-odd weak amplitude between nucleons is approximately seven orders of magnitude smaller than the strong nucleon-nucleon amplitude, \cite{potter1974test, yuan1986measurement, adelberger1985parity}. However, near p-wave neutron-nucleus resonances, interference between nearby s-wave and p-wave amplitudes can produce dramatically enhanced PV effects , \cite{bunakov1981parity, bunakov1983parity}. Asymmetries as large as 10\% have been observed in several heavy nuclei \cite{alfimenkov1983parity}, making these resonances valuable laboratories for studying the weak interaction. 

Two complementary experimental approaches exist for measuring PV in neutron-nucleus resonances. The first method measures the transmission asymmetry of longitudinally polarized neutrons through an unpolarized target
\begin{equation}
    P = \frac{\sigma^+_n(E_n) - \sigma^-_n(E_n)}{\sigma^+_n(E_n) + \sigma^-_n(E_n)},
\end{equation}
where $\sigma^{\pm}_n$ is the (energy-dependent) total cross section for different longitudinally polarized neutrons and $E_{n}$ is the energy of an incident neutron. Observing small neutron transmission differences requires thick targets with areal densities $n \sim 10^{23}$ nuclei/cm$^2$, corresponding to samples approximately 10 cm thick with masses of several kilograms~\cite{seestrom1999apparatus}. This limits the method to nuclei with sufficient available material.  

The second method directly measures the helicity-dependent neutron capture cross section $\sigma_{(n,\gamma)}(E_n)$ (see \cite{masuda1989longitudinal},  \cite{sharapov1991capture} for more details) by measuring asymmetry $P_\gamma$
\begin{equation} \label{eq:Pgamma}
    P_\gamma = \frac{\sigma^+_{(n,\gamma)}(E_n) - \sigma^-_{(n,\gamma)}(E_n)}{\sigma^+_{(n,\gamma)}(E_n) + \sigma^-_{(n,\gamma)}(E_n)},
\end{equation}
where $\sigma^{\pm}_{(n,\gamma)}$ are polarized neutron capture cross sections. This approach enables PV studies using thin, isotopically enriched targets, extending measurements to rare isotopes where large target masses are impractical.

Direct measurement of helicity-dependent neutron capture requires an array of $\gamma$-ray detectors surrounding the target to detect the cascade of 3-4 gamma rays emitted following neutron absorption. 

The small magnitude of PV asymmetries necessitates high statistical precision, requiring large integrated neutron fluxes and correspondingly high event rates in the detector array. When the event rate becomes too high to resolve individual pulses due to detector and electronics deadtime and pulse pileup effects, the detectors must operate in ``current mode" rather than ``pulse counting mode''. In current mode, the signal from the instantaneous particle flux is converted into a continuous electrical current. 


A disadvantage of current mode detection is that most of the information about the relative energies of the events in the detector is lost, and therefore one foregoes the ability to subtract background from the spectrum of detector events in the standard way. In the case of measurements of parity violation, however, since only the weak interaction violates parity, it is possible to design the measurement apparatus to measure spin-dependent asymmetries which isolate the weak interaction process of interest. The effects of background processes usually disappear in the difference of spin state yield, and can often be further characterized well enough in auxiliary measurements to accurately reconstruct the asymmetry of interest. 

The approach of using current-mode $\gamma$-ray detector arrays for PV measurements builds upon extensive prior work. The TRIPLE collaboration at Los Alamos National Laboratory conducted a decade-long campaign measuring PV in neutron-nucleus p-wave resonances using transmission methods. Their instrumentation included a current mode $^{3}$He/$^{4}$He ion chamber for neutron flux measurements~\cite{SZYMANSKI1994564}, a $^{10}$B-loaded liquid scintillator current mode neutron transmission detector~\cite{YEN2000476}, a $^{6}$Li-loaded glass scintillator current mode neutron transmission detector~\cite{BOWMAN1990183}, and a pure CsI current mode gamma detector array~\cite{SEESTROM1999603}. The NPDGamma collaboration designed and operated a CsI(Tl) current mode gamma array~\cite{Gericke:2004xn} for the measurement of parity violation in polarized slow neutron capture on protons. The n$^{3}$He collaboration~\cite{n3He:2020zwd} designed and operated a unique $^{3}$He ion chamber to search for PV in the $n+^{3}$He $\to$ $p+^{3}$H reaction. Searches for PV in the $n+^{4}$He $\to$ n $+^{4}$He,  $n+^{6}$Li $\to$ $^{4}$He $+^{3}$H, and $n+^{10}$B $\to$ $^{4}$He $+^{7}$Li reactions at  the Institute Laue–Langevin (ILL) and the National Institute of Standards and Technology (NIST) employed segmented current mode ion chambers of various types~\cite{PENN2001332, Vesna:2021rma}. 

This paper describes the design, construction, characterization, and testing of a modular array of 24 NaI(Tl) detectors (hereafter, the array) developed by the NOPTREX collaboration for measuring PV through direct detection of neutron-helicity-dependent gamma intensities emitted from the decay of neutron-nucleus resonances. The detector features custom electronics enabling operation in both pulse and current modes. We demonstrate the array's performance through observation of the known 0.7 eV parity-violating resonance in $^{139}$La during measurements at the Los Alamos Neutron Science Center (LANSCE).

\section{NaI Detector Array Design}

The key features of this gamma array follow from the nature of the desired measurements and the environment in which the apparatus operates.

The NaI detector array is designed to operate on a pulsed epithermal neutron beam at the Los Alamos Neutron Science Center (LANSCE), which delivers the neutron beam energies of interest for this work, mainly in the 0.1 eV to 1 keV range. The spallation neutron source operates with a repetition frequency of 20 Hz, and the apparatus is located approximately 20 meters from the neutron production target. The neutron energy is determined using time-of-flight (TOF) from the source to the detector array. When the neutron energy coincides with a resonance in the $A+1$ compound nucleus formed upon neutron capture, the dominant decay mode is through gamma emission. The stable nuclei investigated possess neutron separation energies in the 6-8 MeV range, and the excited $A+1$ nucleus typically emits 3-4 gammas in a cascade to the ground state.

The measurement employs polarized neutrons with a spin flipper located upstream of the target. The target is positioned at the center of the gamma detector array, which measures the total gamma-ray yield as a function of neutron energy and spin state. This configuration enables direct measurement of the helicity-dependent capture cross section asymmetry $P_{\gamma}$, (eq.~\ref{eq:Pgamma}).

These goals and conditions lead to a clear set of technical criteria to be satisfied by the gamma array:

\begin{enumerate}
    \item The gamma array should possess an efficiency which is sufficiently stable over the timescale of the neutron spin flip to avoid false instrumental asymmetries. In particular, the gamma array detectors must be shielded from the magnetic field changes usually employed one way or the other to flip the neutron spin.
    \item The gamma array should absorb as much of the gamma energy emitted by the target consistent with the need for the neutron beam to enter and exit the array. 
    \item The gamma array should be shielded both from external sources of neutrons and gammas as well as from neutron scattered from the target, which can otherwise capture in the gamma detector material and generate background.
    \item The time response of the gamma detectors should be fast enough to introduce negligible broadening in time of the observed resonance and to enable a sufficiently fine statistically-independent sampling of the gamma signal to realize a valid fit to the resonance lineshape.
    \item It should be possible to disable the electronic gain on the detectors during and for a short time after the intense flash of gammas from the proton spallation target that makes the neutrons. It is impractical to bend an epithermal neutron beam out of the line-of-sight of the source using present neutron optics technology. The gamma detectors will therefore inevitably see a bright flash of gammas which can saturate their response and potentially lead to signal distortions and instabilities. 
    \item The gamma array must accommodate a neutron spin transport magnetic field so that the neutron polarization survives not only up to the target but also to the end of the array to enable subsequent measurement to confirm that the neutrons were polarized in the right direction upon absorbing in the target. 
    \item The output signals from the detectors should be chosen to match the relevant capabilities of the analog-to-digital converter used to record the data as a function of neutron time-of-flight.
\end{enumerate}


NaI(Tl) scintillators are well-suited to meet these technical criteria. NaI(Tl) offers high light yield ($\sim$38 photons/keV) and good energy resolution for $\gamma-$rays in the MeV range, making it effective for detecting the gamma-ray cascades following neutron capture, where the total energy release is 6-8 MeV distributed among 3-4 individual gammas. The fast decay time ($\sim$230 ns) is sufficiently short to avoid significant broadening of resonance features in time-of-flight measurements. Large crystal volumes (3$^{''}\times$ 3$^{''}\times$ 5$^{''}$) provide adequate stopping power for gammas with energies of several MeV while maintaining reasonable cost for multi-detector arrays. 

Rexon Components Inc.\footnote{\href{https://www.rexon.com/}{https://www.rexon.com/}} successfully refurbished 24 NaI crystals and encapsulated them in new housings with optical windows for photomultiplier tube coupling. We supplied new PMTs and bases, custom current-mode electronics, data acquisition systems, and the overall mechanical and shielding design. Existing stocks of lead shielding, borated polyethylene, and mu-metal at Indiana University made this detector technology economically feasible for constructing a 24-detector array. Each NaI detector was assembled at Eastern Kentucky University. 

\subsection{Array Geometry and Assembly}

\begin{figure}[!h]
    \centering
    \includegraphics[width=0.4\textwidth]{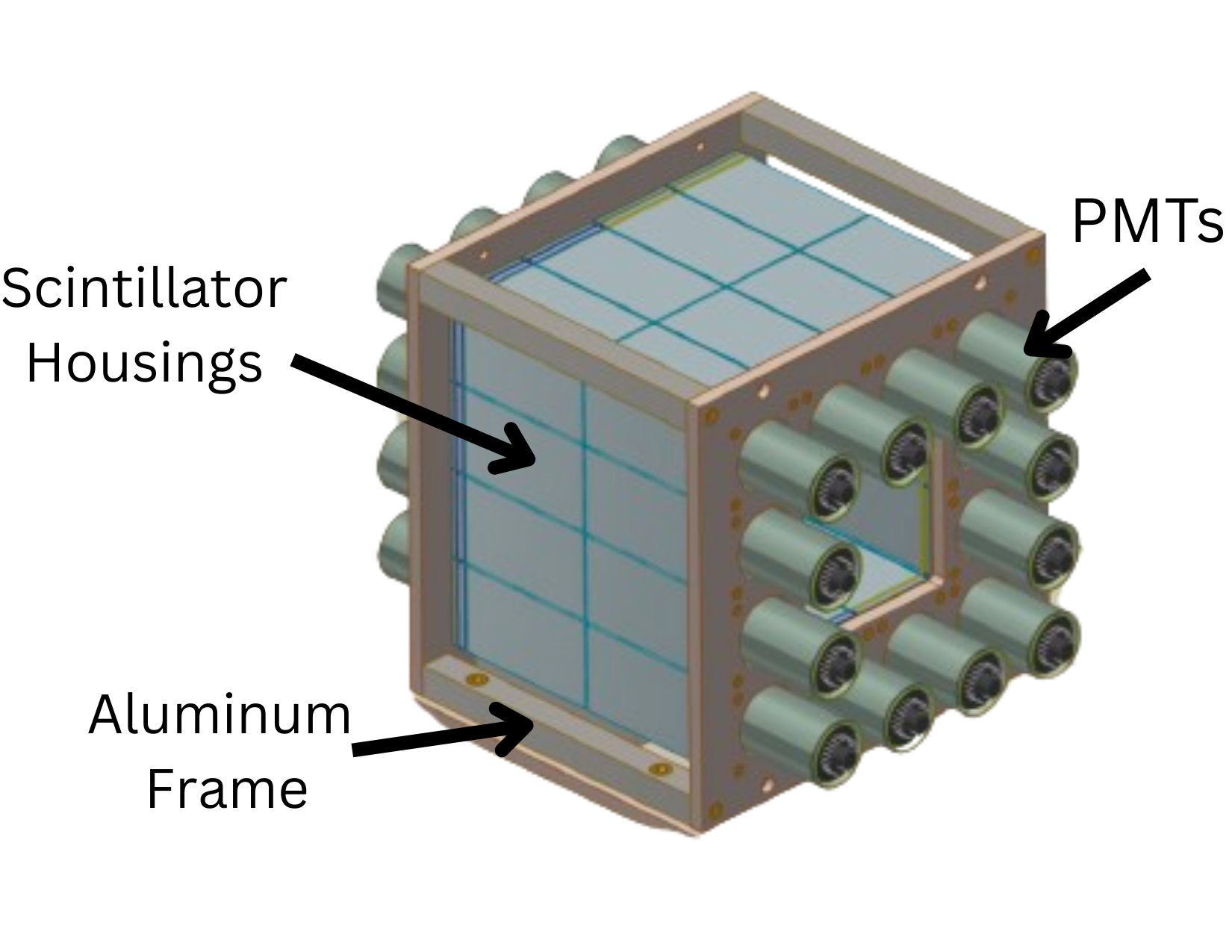}
    \caption{This figure shows the CAD model of the assembled detector array in the aluminum frame machined at IU.}
    \label{fig:NaI_detector_array}
\end{figure}

The array consists of two square rings, with each ring containing 12 NaI(Tl) crystals (see Fig.~\ref{fig:NaI_detector_array}). The NaI(Tl) crystals are placed with crystal ends nearly touching such that the total length of the array is approximately 10$^{''}$, covering approximately 11 steradians (sr) effective solid angle with respect to the center. The array efficiency to detect $\gamma-$rays with energy in the range up to 6 MeV was estimated using MCNP simulation~\cite{kulesza2022mcnp}.
    
\begin{figure}[!htbp]
    \centering
    \begin{minipage}[c]{0.1\textwidth}
        \centering
        \includegraphics[width=\textwidth]{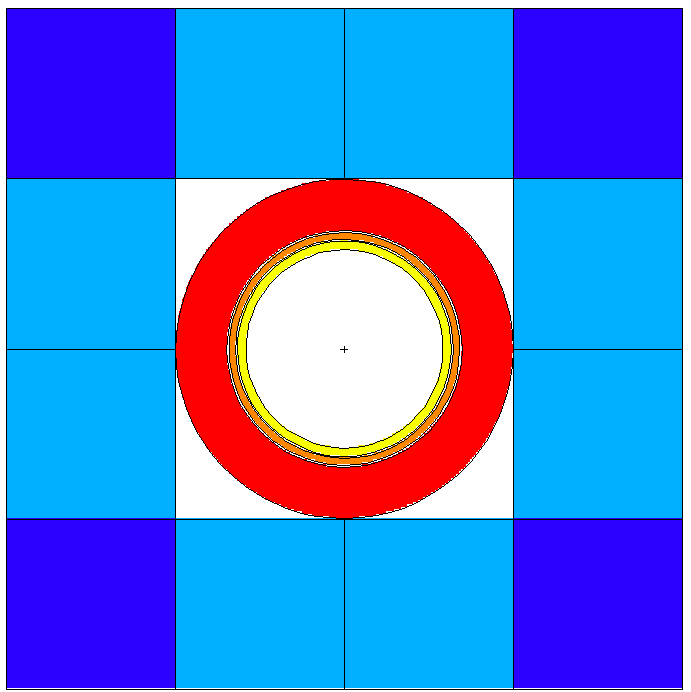}
        \subcaption{(XY) cross-sectional view}
        \label{fig:XY}
        
        \vspace{0.3cm}
        
        \includegraphics[width=\textwidth]{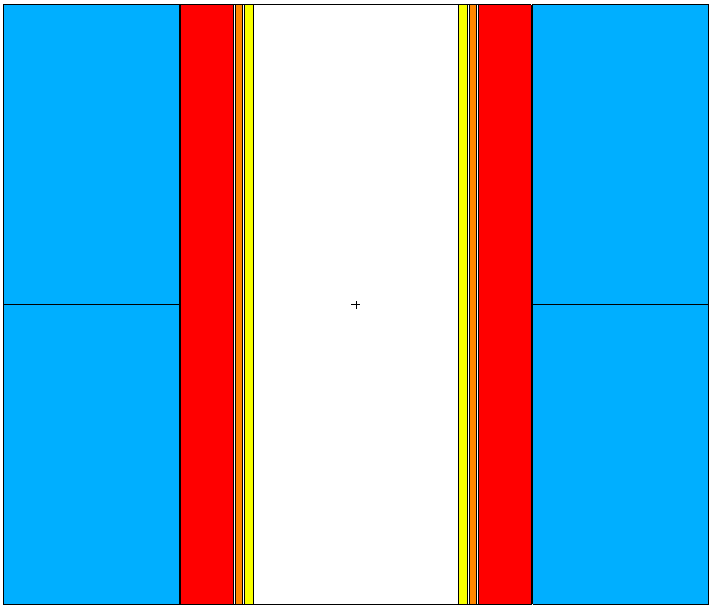}
        \subcaption{(XZ) cross-sectional view}
        \label{fig:XZ}
    \end{minipage}
    \hfill
    \begin{minipage}[c]{0.3\textwidth}
        \centering
        \includegraphics[width=\textwidth]{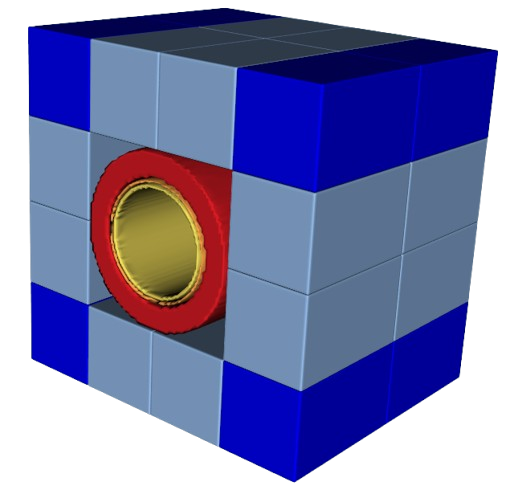}
        \subcaption{3D view of the MCNP model}
        \label{fig:3D}
    \end{minipage}
    \caption{The array MCNP model: NaI(Tl) crystals are shown in light blue (center crystals), and dark blue (corner crystals), Li$_2$CO$_3$-neutron shielding is in red, copper coils are in dark orange, aluminum pipe is in yellow.}
    \label{fig:Array_xsec}
\end{figure}

The efficiency of the array can be expressed as $$\varepsilon_{Array} = \frac{\Omega}{4\pi} \times I\times \varepsilon_{\text{int}},$$ where the array solid angle is $\Omega \approx 11$ sr, $I$ is the fraction of gamma rays entering the array due to attenuation in the shielding materials around the target area, and $\epsilon_{\text{int}}$ is the intrinsic efficiency of the array.

\begin{figure}[!htbp]
    \centering
    \includegraphics[width=0.5\textwidth]{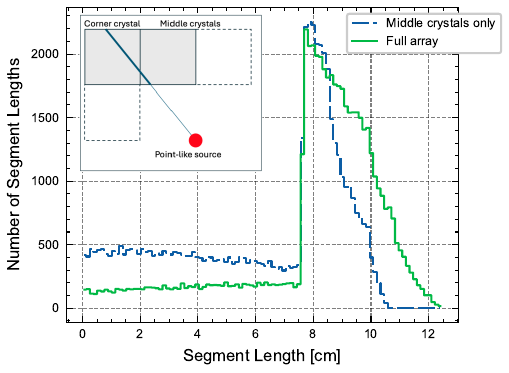}
    \caption{Path length distributions in the center detector alone and in the center detector combined with the corner wedge. Inset: Two middle and corner crystals are shown in light gray. Adding corner crystals increases the gamma ray path length in the array (shown as thick line).}
    \label{fig:Path_length_distr}
\end{figure}

The solid angle covered by the array depends on the length of the array, and the size and geometry of the opening inside the array. Since 24 NaI(Tl) detectors are arranged in two square rings, addition of corner detectors does not increase the array's solid angle. However, they do increase the effective detector depth and, hence, increase the intrinsic efficiency $\epsilon_{\text{int}}$ factor. Distributions of lengths of gamma ray paths inside the center crystal and the center crystal together with corner wedge are shown in Fig.~\ref{fig:Path_length_distr}. The mean length of the gamma ray path increases from 6.2 cm when only center crystals are used, to 8.1 cm when corner crystals are added to the array.

\begin{figure}[h]
    \centering
    \includegraphics[width=0.5\textwidth]{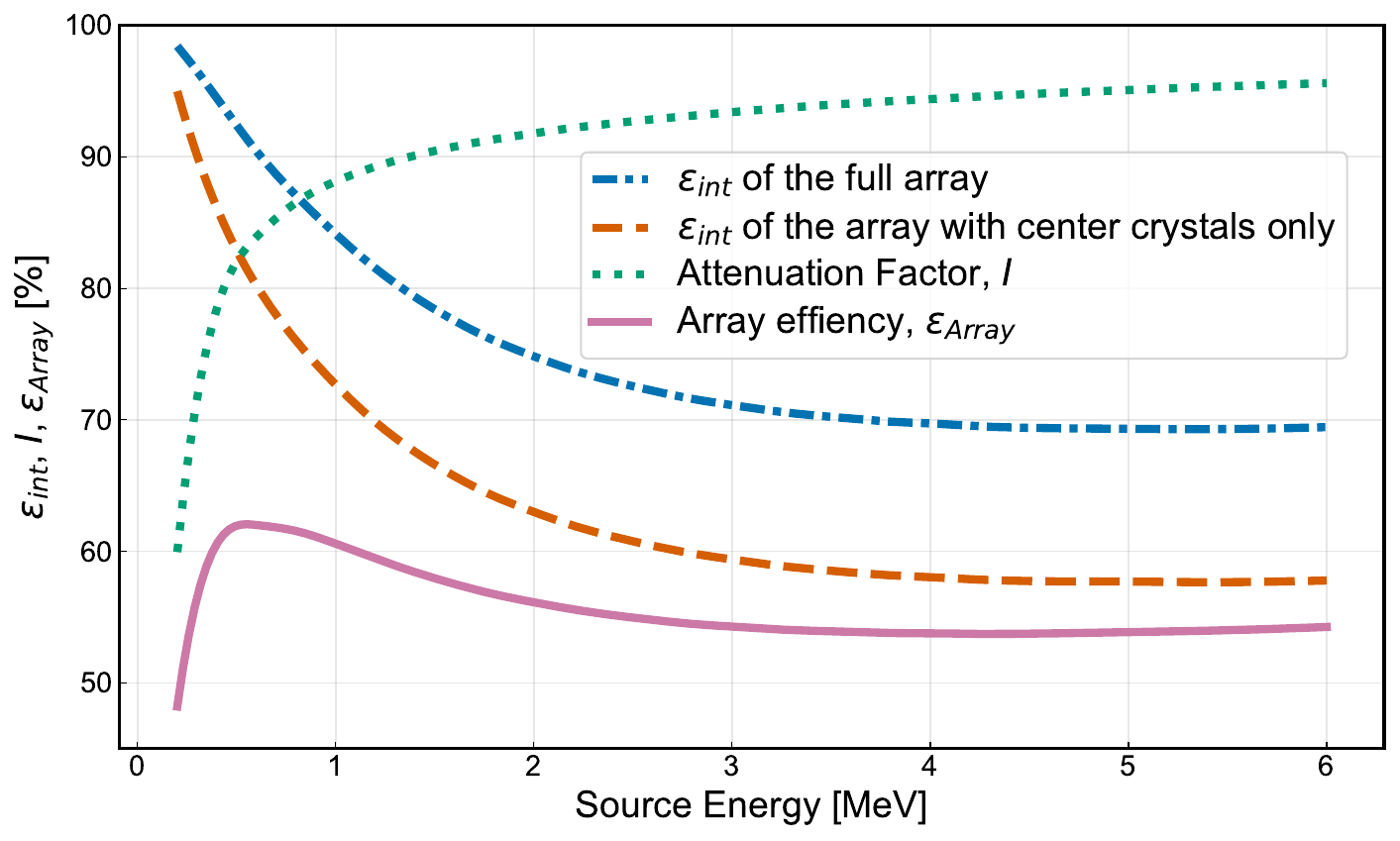}
    \caption{Calculated array's efficiency, $\epsilon_{Array}$, intrinsic efficiency $\varepsilon_{int}$, and attenuation factor $I$}
    \label{fig:Efficincy_I1_graph}
\end{figure}

This observation was confirmed using MCNP simulations. The MCNP model of the array and neutron shielding are shown in Fig.~\ref{fig:Array_xsec}. The MCNP F8 tally was used to accumulate the pulse height spectrum for point-like gamma sources with energies in the range from 0.2 MeV to 6 MeV placed in the geometrical center of the array. Simulations were conducted for the array's configurations using only center crystals (center crystals are shown in light blue in Fig.~\ref{fig:Array_xsec}) and when all crystals were included. Since the array is used in current-mode, the intrinsic efficiency was calculated based on the total counts in the observed gamma spectrum. The results are shown in Fig.~\ref{fig:Efficincy_I1_graph}. Addition of NaI(Tl) crystals to the corners of the array increases the overall intrinsic efficiency, $\epsilon_{int}$, by 12\% at 2 MeV to approximately 75\%. 


\begin{figure}[!h]
    \centering
    \includegraphics[width=0.45\textwidth]{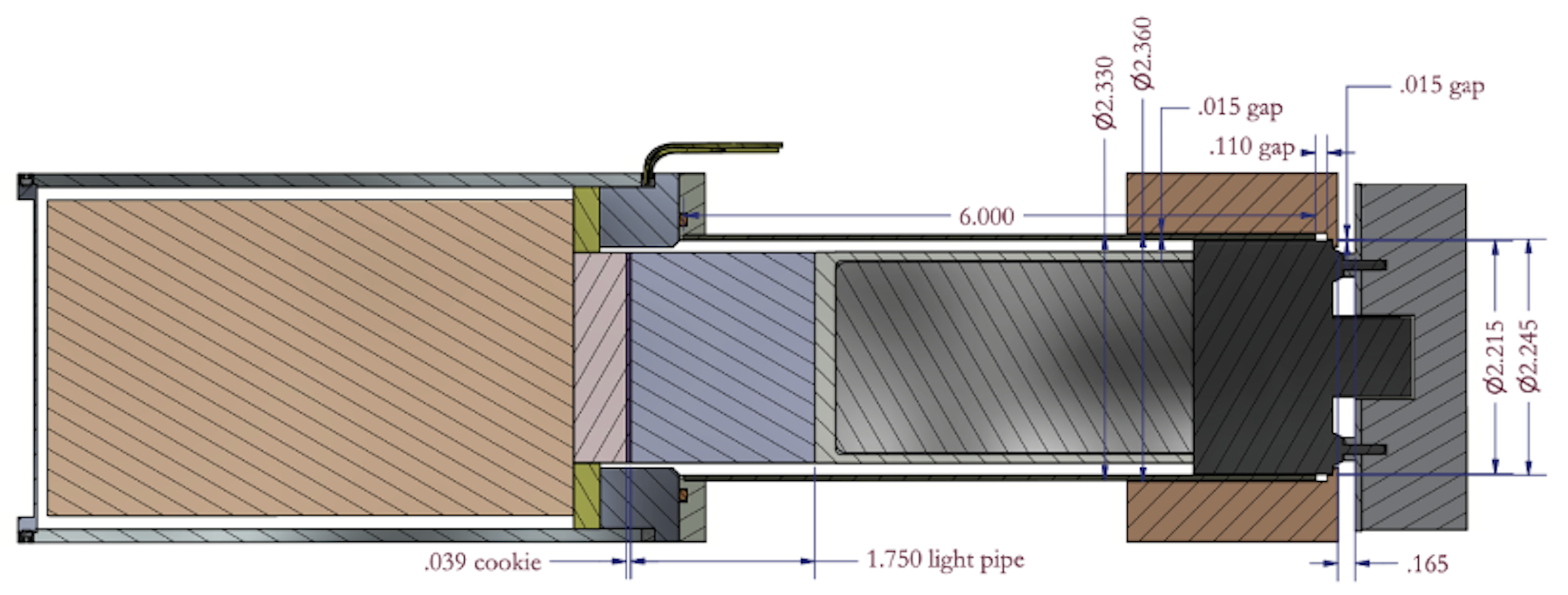} 
    \caption{CAD drawing of the NaI detector housing and component parts.}
    \label{fig:detector_housing}
\end{figure}

Finally, the effect of shielding and copper coils was estimated using MCNP simulation. The flux through the array was normalized using the flux from the 6-MeV gamma point source when no shielding was placed in the setup. The attenuation factor $I$ as a function of the source energy is shown in Fig.~\ref{fig:Efficincy_I1_graph}. 
Using the obtained results, the overall array efficiency was calculated for gamma rays energy from 0.2 MeV to 6 MeV (see Fig.~\ref{fig:Efficincy_I1_graph}).


\subsection{Detector Enclosure and Shielding Design}

The housing for the detectors was designed to minimize exposure to external light and magnetic fields, which can affect detector performance. The detector housing uses a mu metal magnetic shield made of a nickel-iron alloy. These recycled shields served as a constraint to the housing design. To prevent light leaks and to hold the shield in place, the detector housing consists of an aluminum magnetic shield flange ring that is bolted onto the top of the aluminum scintillator housing over a rubber o-ring gasket. The magnetic shield is press-fitted into this flange. A Delrin\textsuperscript{\textregistered} clamp-ring was made to slide onto the top of the shield, with the pins of the PMT sticking out to connect to electronics. The Delrin\textsuperscript{\textregistered} clamp-ring is held down by a set of four nonmagnetic extension springs that hook onto nylon eyebolts that are screwed into the clamp-ring and the magnetic-shield flange (see Section \ref{sec:detedesign} for more details). 
These springs provide constant tension that keeps the clamp-ring, shield, and PMT in place. To further prevent light leaks, the joints between the flange, shield, and clamp-ring are sealed with a rubber gasket compound. The parts for this housing were designed by Jack Doskow and John Vanderwerp at Indiana University, and machined at Indiana University and the University of Kentucky (Fig.~\ref{fig:detector_housing}). 

The array stand was designed out of 80/20 extruded aluminum beams in an effort to reduce stray magnetic fields. The stand design allows for a considerable amount of shielding on all sides of the array, and is capable of supporting the weight of multiple layers of lead bricks.

\subsection{PMT Placement and Efficiency}

Each of the NaI(Tl) detectors was coupled to a Hamamatsu R550 PMT \footnote{\href{https://www.hamamatsu.com/us/en/product/optical-sensors/pmt/pmt_tube-alone/head-on-type/R550.html}{https://www.hamamatsu.com/us/en/product/optical-sensors/pmt/pmt\_tube-alone/head-on-type/R550.html}}, with a custom electronics board from Indiana University for voltage division and preamplification (Fig.~\ref{fig:electronics}). This custom electronics board allows for the R550 PMT, designed to operate at positive high voltage, to be used at negative high voltage. To enable negative high voltage operation a wire connected a high voltage pin to the glass exterior of the PMT. This wire ran through a groove carved into the side of the plastic PMT cap and was held in place by a thin layer of 2-part epoxy. 

Because the exterior of the PMT is held at high voltage, it was necessary to electrically separate the PMT from the aluminum scintillator housing. For this purpose, we used a transparent plastic lightguide between the PMT and scintillator. The lightguide was coupled to the PMT using a 2-part optical epoxy, and the coupled PMT-lightguide assembly was coupled to the scintillator using a 1-mm-thick silicon disk. This soft silicon disk is slightly compressible, ensuring adequate optical contact when compressed by the force of the springs in the shield housing (see Section \ref{sec:detedesign}).

\subsubsection{Lightguide Length and Magnetic Shield Testing}

In order to determine the optimal length for the lightguide, it was necessary to consider the effects of lightguide length on the efficacy of the magnetic shield. Because the photoelectrons within the PMT would be most sensitive to external magnetic fields near the photocathode-end, this section of the PMT needed to be located well within the boundaries of the mu metal shielding without causing the opposite end to be unprotected. Several measurements were performed to determine the optimal position of the photocathode within the shield, and therefore the best length for the lightguide. 

\begin{figure*}[!ht]
\centering
\includegraphics[width=12cm, height=4.8cm]{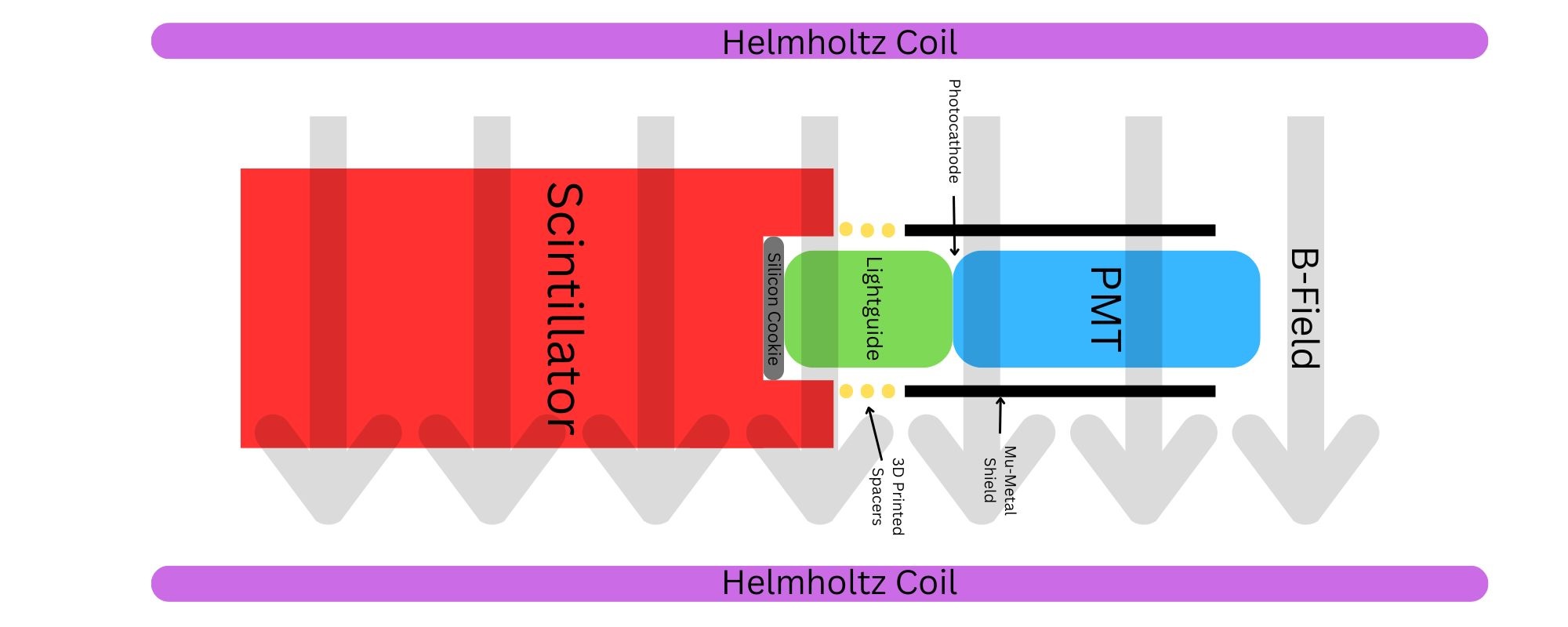}
\caption{The design of the magnetic shield position tests.}
\label{fig:magshieldtest}
\end{figure*}

In these experiments, 1/4 inch spacers created by 3D printing  were used to change the position of the mu metal shield relative to the photocathode. This allowed for the use of one polished lightguide of a single size, rather than multiple different lengths of cut/polished lightguide. For each shield position, the NaI detector was placed in a Helmholtz coil in the transverse direction, with the photocathode near the center where the field should be the most uniform, and a $^{60}$Co spectrum was collected over the course of 10 minutes for a positive current, negative current, and zero current in the coil (Fig.~\ref{fig:magshieldtest}). The intensity of the field was monitored using an Adafruit MLX90393\footnote{\href{https://www.adafruit.com/product/4022}{https://www.adafruit.com/product/4022}} triple-axis magnetometer, which was readout using Python code on a Raspberry Pi. With an external power supply supplying a current of 8 amps, a field strength of roughly 15 Gauss was achieved near the photocathode.

A Red Pitaya board\footnote{\href{https://redpitaya.com}{https://redpitaya.com}} flashed with a trapezoidal filter multichannel pulse height analysis firmware was used for data acquisition. For each of the $^{60}$Co spectra that were collected, we used a Gaussian fitting function in Python to determine the location of the photopeaks. The relative heights of the photopeaks were used as an indication of the light-collection efficiency of the detectors. This efficiency indicated how much the external magnetic field affected each detector based on its photocathode location within the shield.

After spectra were recorded for the positive, negative, and zero magnet currents for each shield arrangement and the locations of the photopeaks were noted for each spectra, the average differences between the photopeak heights with the field on and off were determined. This average difference represented the amount of protection provided by the shield in that arrangement. The results of this experiment showed a clear minimum for the magnetic field influence. This result confirms that on one side of the minimum, the photocathode end of the PMT was too close to the opening of the shield while on the other side of the minimum the anode end of the PMT was exposed, which caused another increase in magnetic field influence. In between these areas of exposure, we found the optimal arrangement to protect both the photocathode and the anode was to align the bottom of the shield 1" below the photocathode end. Because of these results, in addition to other constraints of our design, it was decided to use a lightguide length of 0.75" for the detectors.

\subsection{Detector Electronics and Pre-amplifier}

\begin{figure}[!h]
    \centering
    \includegraphics[clip, scale = 0.5]{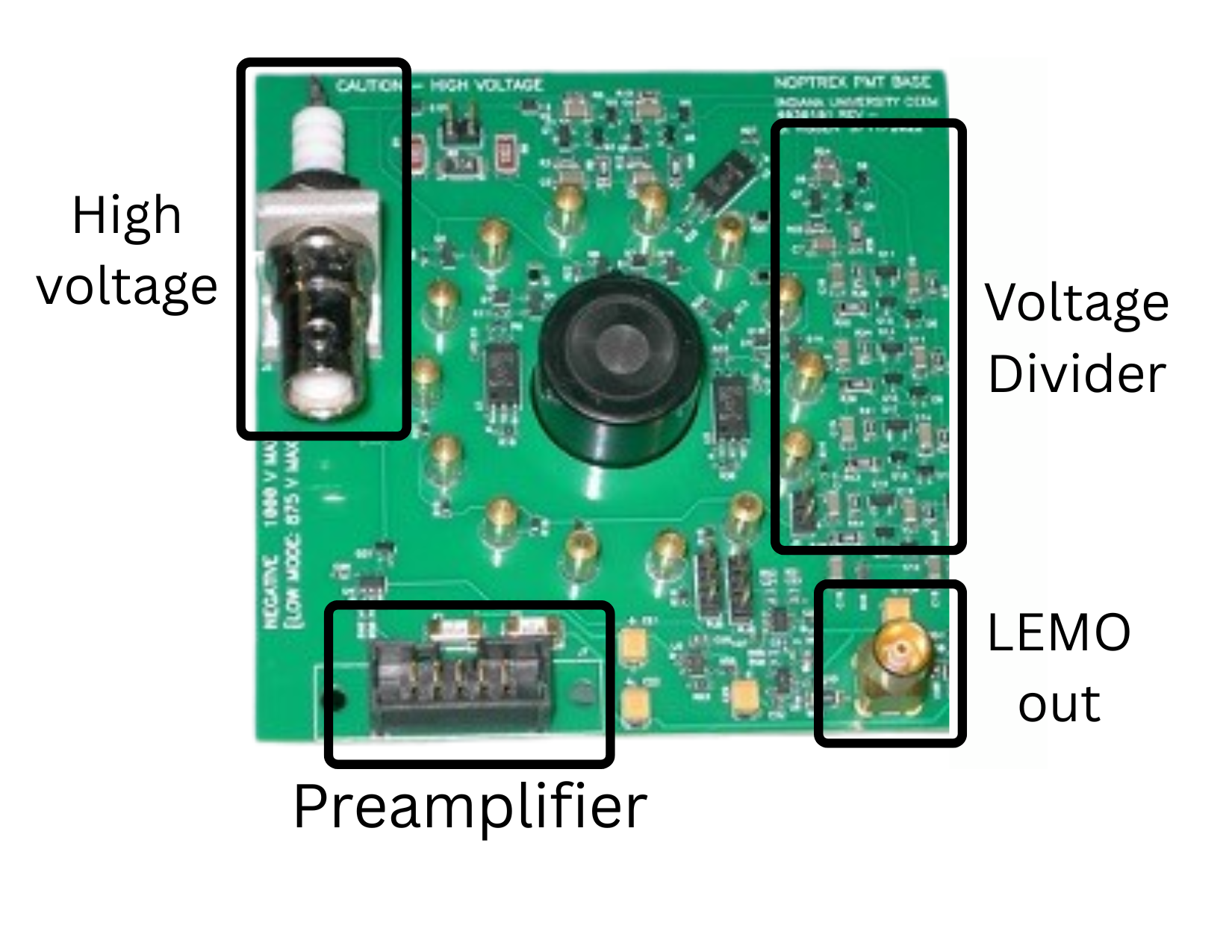}
    \caption{Custom voltage-division and preamplification electronics designed at Indiana University.}
    \label{fig:electronics}
\end{figure}

In order for the detectors to be operated in both current (integrating) mode and pulsed (counting) mode, a custom electronics board and preamplifier was created as outlined in figure~\ref{fig:electronics}. The high voltage supply delivers negative high voltage to the PMT through voltage divider network. The preamplifier is powered through $\pm$6 V and ground. The LEMO out of the board is matched to the 50 $\Omega$ load to the digitizer. Jumper pins on the board allow for the operation in either current or pulsed mode. 

The circuit~\cite{PMT_base_circuitdiagram} was developed to match various specifications of both the PMT and the digitizer (see section~\ref{sec:JPARCDAQ} for more details). The overall gain of the preamplifier was matched to the 1.5 V dynamic range of the digitizer. Additional jumper pins on the board allow for a low and high gain to accommodate different expected fluxes from different neutron sources. An additional feature that was included in the design was the ability to blank the gain of the board during the initial gamma flash of the pulsed neutron source. When protons initially hit the target, a burst of gammas emit from the target and can saturate the detectors. To prevent this, the gain in the early dynodes can be blanked during this period so that the PMTs do not get saturated.

The power dissipation per board is about 2 W, or 48 W total for the array. Since the array is enclosed in lead shielding (see section~\ref{sec:arraySetup}), an air cooling loop with fans was installed to take away the heat generated by the boards. 

\section{Detector Assembly}
\label{sec:detedesign}
The process of designing, constructing, testing, and characterizing the NaI(Tl) detectors was conducted primarily by undergraduate research assistants at Eastern Kentucky University, with the assistance of researchers from Indiana University and the University of Kentucky. Students at Eastern Kentucky University were able to use this experiment to gain a unique insight to the dynamics of nuclear physics research and instrumentation, while also gaining hands on skills and performing analysis on the data collected during detector testing and characterization. 

\subsection{PMT-Lightguide Coupling}
To couple the lightguides to the PMTs, the RTV 615 Silicone potting compound was used. The compound is a two-part mix and is optically transparent in the expected wavelength spectrum of the PMT. 

For the coupling setup, a mu-metal shield and PMT clamp ring were used as a stand to hold the PMT upright, and the PMTs were inserted, upside down, into the shield and rested on a scrap light guide. This allowed the photocathode-end to stick out of the shield and be accessible for cleaning and coupling. To ensure consistent results, the PMTs were only coupled at a rate of 5-6 at a time. To prepare for coupling, the PMTs were cleaned with isoporpyl alcohol and allowed to dry. The silicone potting compound was then applied, and a heat gun was used to remove bubbles that remained from the mixing.  

After applying the compound to the PMT, the light guide was placed onto the PMT surface and a generous amount of downwards pressure was applied to it. This pressure caused the compound to fill the gap, and after consistently applying pressure, it was noticeable that the residual air bubbles were being pushed out the sides of the joint. By strategically applying pressure and gently moving the light guide, it was possible to remove most of the residual bubbles. Electrical tape was then wrapped around the joint to hold the lightguide in place, and the coupled PMT was inserted into the shield with the lightguide-end down, so as to use the weight of the PMT to apply pressure on the coupled joint (see Fig.~\ref{fig:taping}a). Following the coupling process, the compound joint was then allowed to set for multiple days, or up to a week, before the taping process. 

\subsection{PMT Taping}

For the taping of the coupled lightguide/PMTs, we used 3 kinds of tape: RF EMI shielding tape (conductive tape), Kapton\textsuperscript{\textregistered}  tape, and 3M-23 Scotch\textsuperscript{\textregistered} self-bonding electrical tape. 

\begin{figure}[!h]
    \centering
     \includegraphics[clip, scale = 0.30]{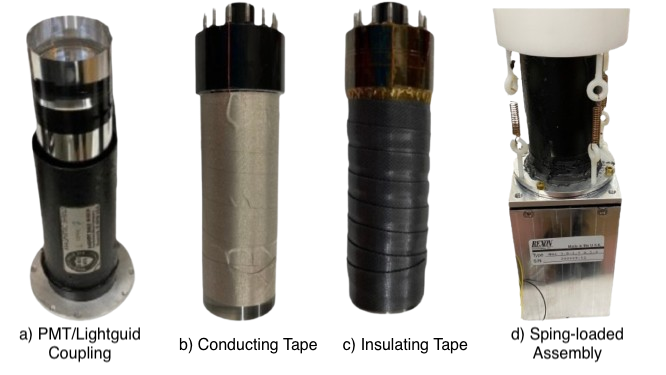}
    \caption{After coupling, the light guide/PMT pairs were placed upside down into shields to hold them until the coupling compound cured. Tape was also applied around the joint to prevent the light guides from moving before the compound cured. The detectors were then wrapped in a layer of conductive and insulative tape.}
    \label{fig:taping}
\end{figure}

The first layer of taping was the conductive tape. The conductive tape was applied over the wire leading from the PMT pin, and wrapped continuously, with a slight overlap, to the end of the PMT, with about a 1/3'' clearance on each end (see Fig.~\ref{fig:taping}b). This clearance was to ensure that the conductive tape was not exposed after being wrapped with insulating tape. To ensure a proper connection between the PMT pin and the insulating tape, the resistance between the two was measured using a digital multi-meter. This resistance measurement was performed across the tip of the PMT pin attached to the wire, and a point on the conducting tape at the far end of the PMT. Any resistance reading near or less than 1\textOmega \ was considered satisfactory.

%
%

The layer of conductive tape was then enclosed in a layer of insulating 3M-23 Scotch\textsuperscript{\textregistered} self-bonding electrical tape. Unlike the conductive tape, which was given a clearance of 1/3'' from each end of the PMT, the electrical tape was applied up to each end so as to ensure a complete coverage of the conductive tape. This layer was applied as thinly as possible, while also creating a slight overlap in each rotation due to the self-bonding nature of the tape (see Fig.~\ref{fig:taping}c). In addition to this electrical tape, insulating Kapton\textsuperscript{\textregistered} tape was also applied to the plastic end cap of the PMT, around the wire, to reduce the likelihood of the wire being exposed during assembly. Two layers of Kapton\textsuperscript{\textregistered} tape were applied, with a slight overhang on either end of the plastic PMT cap. Photos were recorded of each PMT throughout the stages of taping, in addition to the resistance reading between the pin and conductive tape, for future troubleshooting needs. After taping, each PMT-lightguide combination was inserted into a magnetic shield, and locked into the spring-loaded housing assembly (see Fig.~\ref{fig:taping}d), which was then bolted into the top of top of the scintillator housing. A thin silicon wafer was placed between the lightguide and the optical surface in the scintillator housing.

\section{Detector Performance}

\begin{figure}[!h]
\centering
\includegraphics[width=1.0\linewidth]{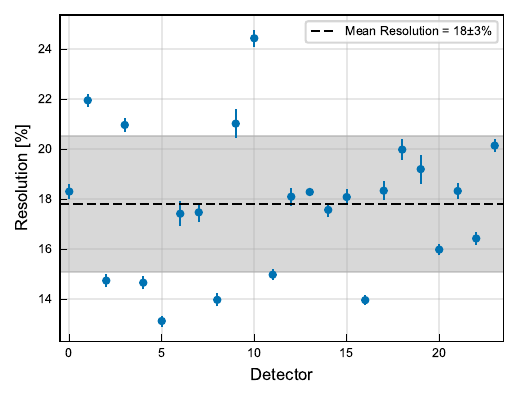}
\caption{Energy resolution of each NaI detector measured at the 662 keV photopeak of $^{137}$Cs.}
\label{fig:Energy_resolution_661leV}
\end{figure}

Before assembling the array, we conducted tests to determine the energy resolution of the detectors. To determine the energy resolution of the detectors, we collected a $^{137}$Cs spectrum for each detector, and measured the full width at half maximum value of the 662 keV photopeak. These resolution measurements were compared to the measurements provided by the manufacturer for each scintillator crystal. We collected these spectra for each detector using the same electronics board for amplification and voltage division. In addition, we also collected spectra using a single detector for each electronics board to ensure the proper performance of the boards. Once we had collected and analyzed these spectra and ensured the quality of the detectors, we were able to continue in assembling the detector array as intended. See Fig.~\ref{fig:Energy_resolution_661leV} for the FWHM performance of all detectors, which is defined as the FWHM of the photopeak divided by the photopeak mean. 

\section{First JPARC Test Run}

\subsection{Experimental Setup}

\begin{figure*}[!ht]
\centering
\includegraphics[width=0.8\linewidth]{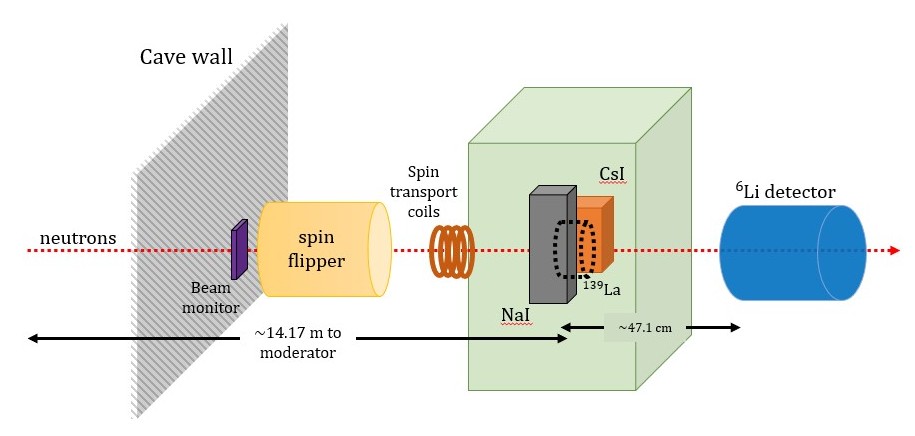}
\caption{Diagram of the experimental setup in BL 10.}
\label{fig:temp2}
\end{figure*}

\begin{figure}[!ht]
\centering
\includegraphics[width=0.45\textwidth]{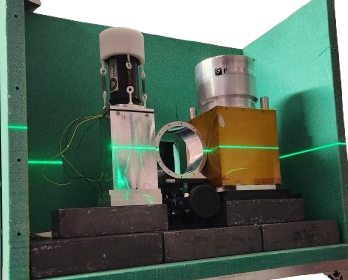}
\caption{A NaI (left) and CsI (right) detector placed on either side of the La target holder, inside of the lead and borated polyethylene shielding.}
\caption{A NaI (left) and CsI (right) detector placed on either side of the La target holder, inside of the lead and borated polyethylene shielding.}
\label{fig:JPARC_g_detectors}
\end{figure}

Before use at LANSCE, two detectors were sent to the Japan Proton Accelerator Research Complex (J-PARC) Material Science Division, Beam Line 10 (BL 10), to test their performance in a neutron beam. The 1 MW, 25 Hz beam was used with a 670 g $^{139}$La target to characterize the detectors with the known $^{139}$La spectrum, particularly in the s- and p-wave region of interest (see Fig.~\ref{fig:temp2}). To compare to the NaI detector, a CsI gamma detector was also placed next to the La target (Fig. \ref{fig:JPARC_g_detectors}), as well as a $^{6}$Li transmission detector downstream from the target. An ex-situ $^{3}$He spin flipper was also placed in the neutron beam to allow for future asymmetry analysis.

\subsection{JPARC Data Acquisition System (DAQ)}
\label{sec:JPARCDAQ}

To manage the DAQ system for the NOPTREX NaI detector array in the test run at JPARC, we used the CAEN DT5560SE\footnote{\href{https://www.caen.it/products/dt5560se/}{https://www.caen.it/products/dt5560se/}} open FPGA digitizer featuring 32 analog input channels, 14- bit ADC, and 125 MS/s processing capabilities. The board defaults with a precompiled multichannel pulse height analysis firmware that includes an oscilloscope and energy spectrum. 

For the JPARC test-run DAQ system, firmware allowing for a decimation of the standard board data rate of 125 MS/s was required. In CAEN's firmware program SciCompiler, we designed firmware which imported the analog data for each neutron pulse and digitized the data at a custom rate with the built in oscilloscope module. The oscilloscope module performed an off-board decimation in the software, and the decimation rate could be set in the software.  For the JPARC test-run, the oscilloscope module sampled the signal at a frequency of 1.95 MS/s, which corresponds to a decimation rate of 64. Using this decimation, we collected roughly 10,000 samples (or time bins) over the course of 5.12 ms for each neutron pulse. This range was chosen because we did not need to analyze the entire duration of the neutron pulse. The chosen range focused on the 0.7 eV p-wave $^{139}$La resonance. The oscilloscope also performed the data readout to a PC. This process was triggered for each new neutron pulse by the accelerator T$_0$ signal provided by the facility. 

Software was written to take in a configuration file that contains parameters for the software and firmware, like the output file path, trigger threshold, and decimation factor, and export the data to a ROOT file. It used two interface classes: one that interacts with the firmware and one that handles file creation. Inside the configuration file, we provided the total number of events we desired for the DAQ run as well as the number of events per file. The software would loop until the total number of events had been recorded but would write to a new file periodically with a smaller number of events. This was done to ensure that any crashes wouldn't cause the loss of a significant amount of data.

Using a parameter in the configuration file, it was also possible to set the number of values stored per event. This can be used to reduce the amount of memory required to store the data since the firmware stores 16,384 entries per event.

In order to test the performance of the firmware and software without access to the running beamline, we used raw data from a previous experiment conducted by the NOPTREX collaboration at LANSCE. This data was generated by a waveform generator, in addition to a 20 Hz rectangular pulse to mimic the T$_0$ trigger. This allowed us to collect a series of decimated waveforms to test the functionality and timing of the firmware and software that mimicked the expected signal and timing.

\subsection{Data Analysis of s- and p- Wave Resonances}

As stated previously, for each neutron pulse the signals in the NaI detector were integrated every 512 ns (which is designated as one time bin) for a total of 5.12 ms giving 10,000 time bins. The firmware was programmed to allow for a pre-trigger of 1,000 time bins for baseline subtraction. Each pulse was baseline subtracted and a simple threshold trigger was used to find the T$_0$ from the gamma flash. Once the data was baseline subtracted and zeroed from the gamma flash, the conversion from time bin to time-of-flight from the T$_{0}$ was performed through known resonances from calibration foils and verified by the known distance between the spallation source and La target. For each run, all the neutron pulses were integrated, which resulted in the TOF spectrum as shown below in Fig.~\ref{fig:jparc2}. The y-axis is arbitrary integrated ADC counts from the DAQ. 

    \begin{figure}[!h]
    \includegraphics[width=1.0\linewidth]{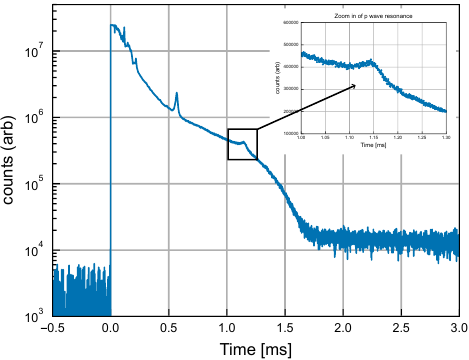}
    \caption{Neutron time-of-flight spectrum from NaI(Tl) over many events in current mode showing the s- and p-wave resonances of $^{139}$La. }
    \label{fig:jparc2}
\end{figure}

Using the conversion between energy and time-of-flight via 
\(
E  = \frac{1}{2} m \left(\frac{x}{t-T_0}\right)^2
\)
the 0.7 eV p-wave resonance should correspond to 1.14 ms, which agrees well with the data. Also note the s-wave resonance peak at about 0.55 ms, higher energy resonances at shorter times, and background above 1.5 ms. 

\section{LANSCE Parity-Violation Experiment}

\begin{figure*}[!t]
\centering
\includegraphics[width=0.8\linewidth]{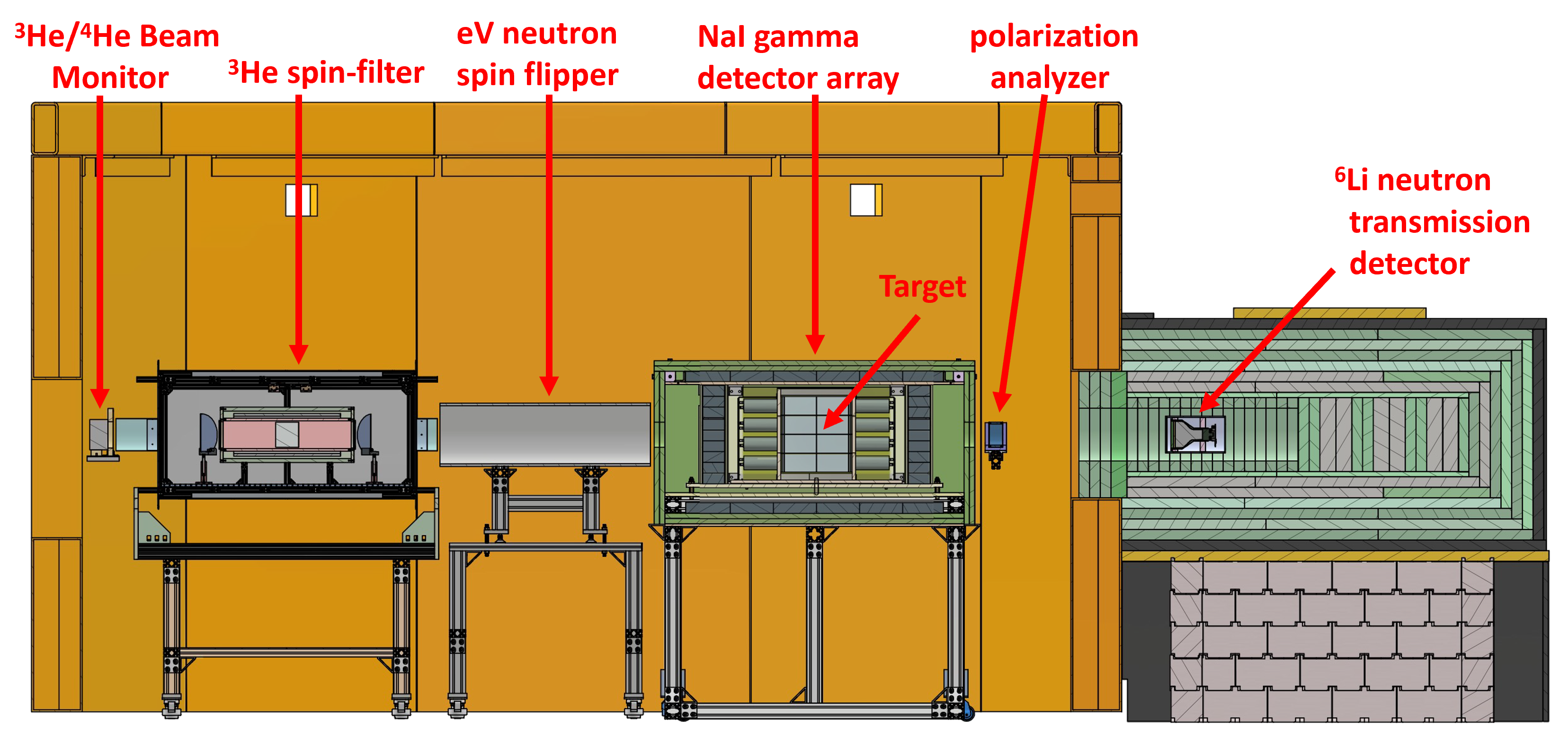}
\caption{Experimental design of the NOPTREX PV resonance search at LANSCE.}
\label{fig:fp12}
\end{figure*}

\begin{figure*}[!t]
\centering
\includegraphics[width=0.45\linewidth]{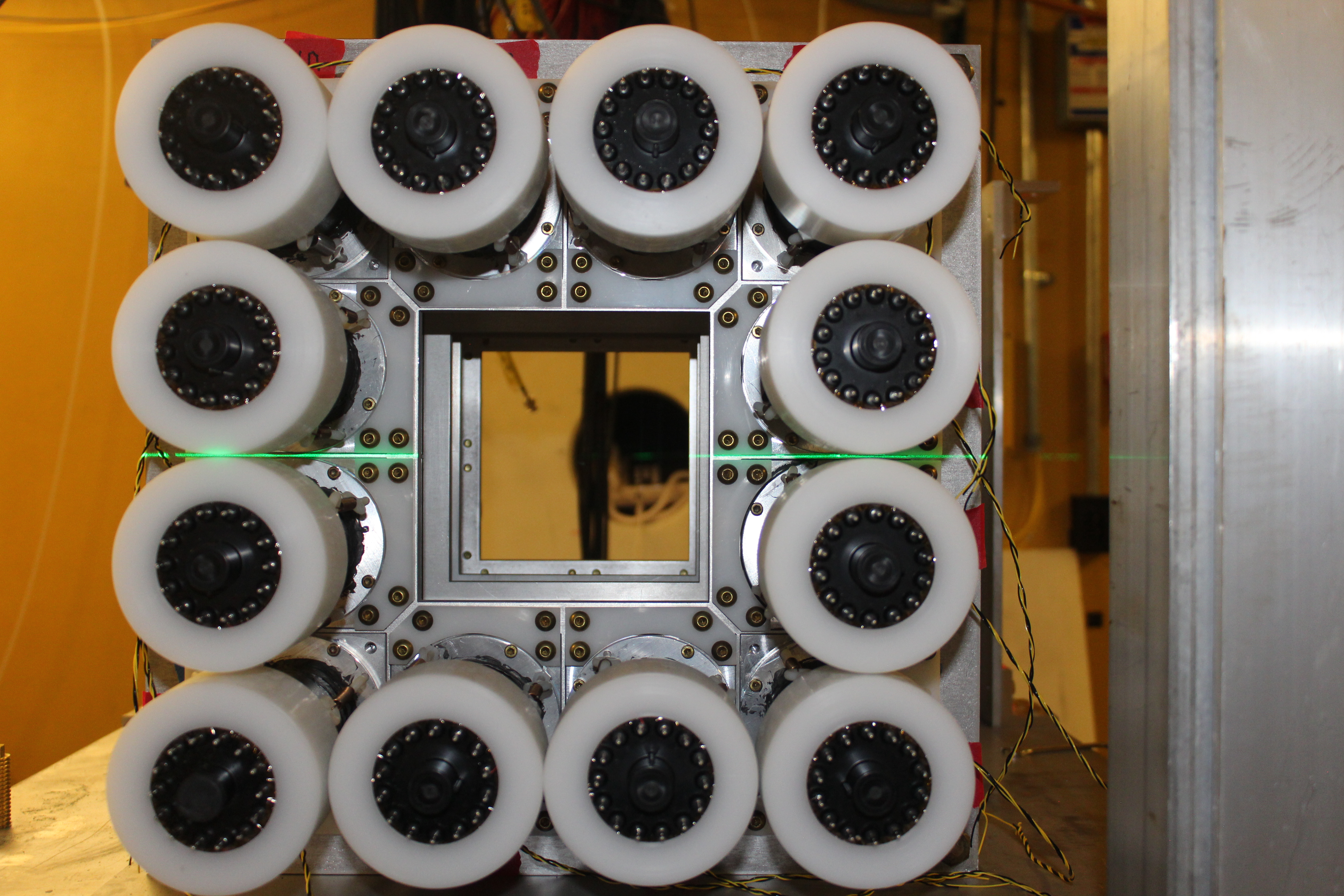 }
\includegraphics[width=0.45\linewidth]{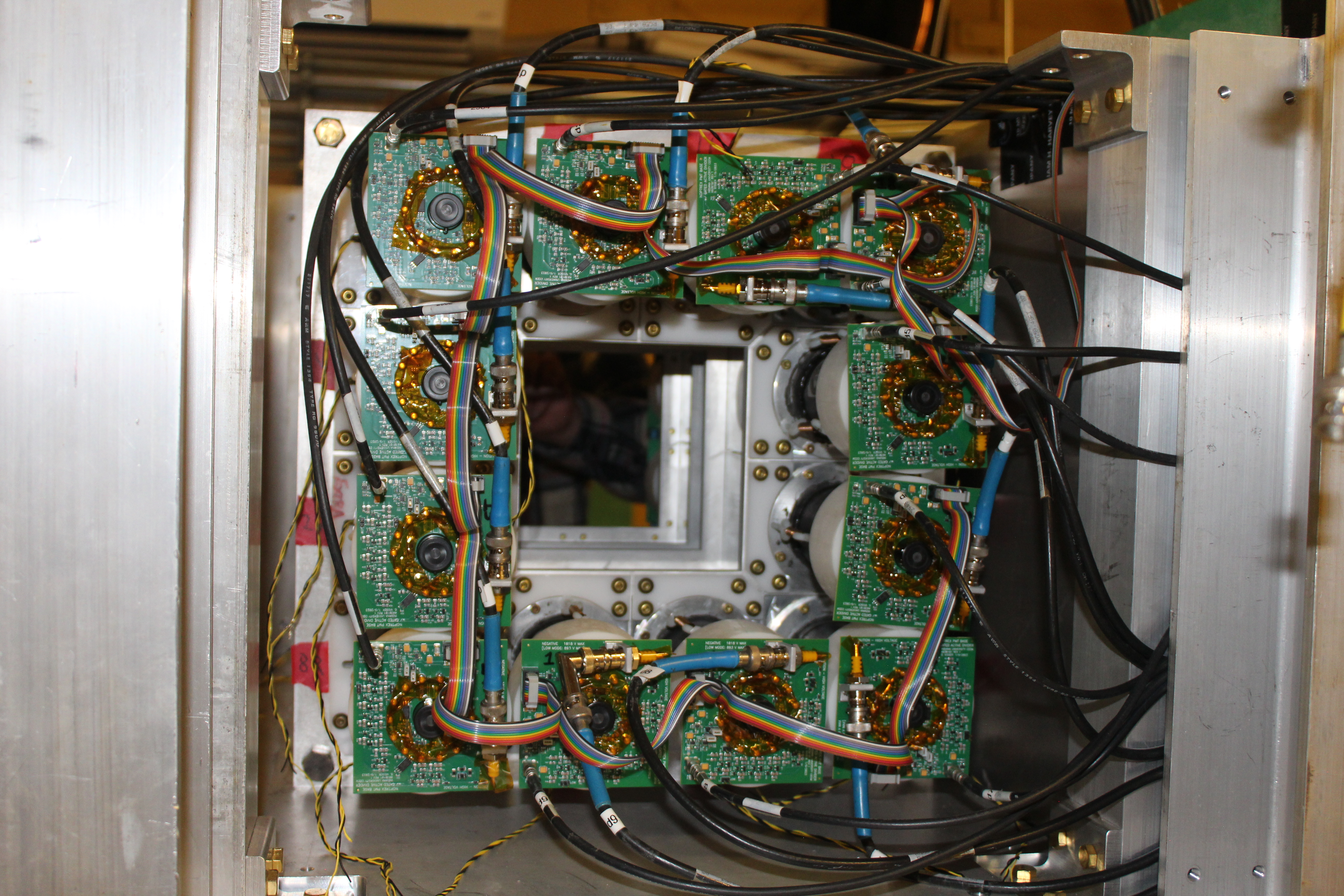}
\caption{Upper: One face of the detector assembly, prior to attaching the PCBs. Lower: One face of the detector assembly after attaching the PCBs and connecting the cabling.}
\label{fig:Detector arrangement photo}
\end{figure*}

\subsection{The LANSCE Facility}
The Los Alamos Neutron Science Center (LANSCE) at Los Alamos National Lab (LANL) provides a 20 Hz, unpolarized, pulsed neutron beam. Detector array testing and operation was done on flight path 12 (FP12) for this experiment due to its open, customizable space. Data was taken with a 32 channel CAEN DT5740\footnote{\href{https://www.caen.it/products/dt5740/}{https://www.caen.it/products/dt5740/}}, triggered by a TTL facility trigger.


\subsection{Experiment Summary}

The full detector array was used for the first time by the NOPTREX collaboration at LANSCE in a search for new parity-violating resonances in various rare-earth elements. These new parity violating resonances could serve as potential candidates for a parity- and time-reversal-odd measurement. To measure any new PV asymmetries, we developed multiple components in addition to the NaI(Tl) including major components: a $^{3}$He neutron spin-filter to longitudinally polarized neutrons and an eV neutron spin flipper to flip the neutron spin by 180 degrees (Fig. \ref{fig:fp12}). The operating principles of these devices have been described in an earlier paper~\cite{Schaper2020}. 

This run started with measuring the known asymmetry of $^{139}$La as a benchmark of the performance of the apparatus. The La target was then replaced with various other rare-earth targets of natural isotopic abundance, including Pr, Tb, Tm, Ho, and Yb.


\subsection{Array Setup}
\label{sec:arraySetup}


All detectors were gain-matched using a $^{137}$Cs radioactive source. The PMT's standard operating voltage of -776 V was used as a baseline, and the DAQ bin location of the 662 keV $^{137}$Cs photopeak was matched in each detector (shown previously in Fig. \ref{fig:Energy_resolution_661leV}). Operating voltages at the end of the gain-matching procedure ranged from -675 V to -973 V. The range of high voltages needed to get similar detector responses is not surprising in view of the many sources of possible detector-detector differences in crystal light output, light reflection from the diffuse reflecting surfaces inside the crystal, light transmission through the ``cookies" used to couple the light into the PMT window, efficiency of light transport through the PMT windows, and light response differences of the different PMTs. The twenty-four detectors were then placed in two sets of twelve, one set downstream and one set upstream of the target center location. Fig.~\ref{fig:Detector arrangement photo} shows the physical assembly of one face of the array with and without the electronic boards and wiring.

\begin{figure*}[!h]
\centering
\includegraphics[width=0.45\textwidth]{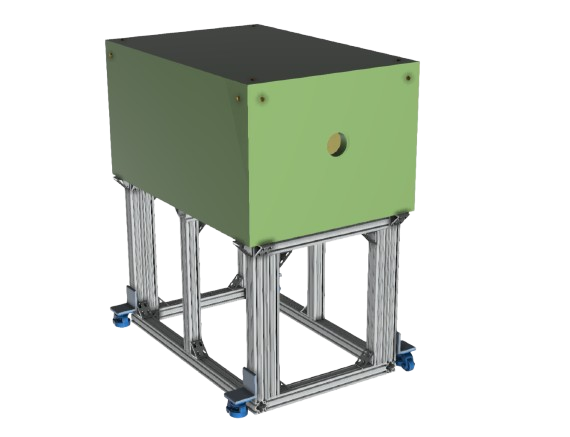}
\includegraphics[width=0.45\textwidth]{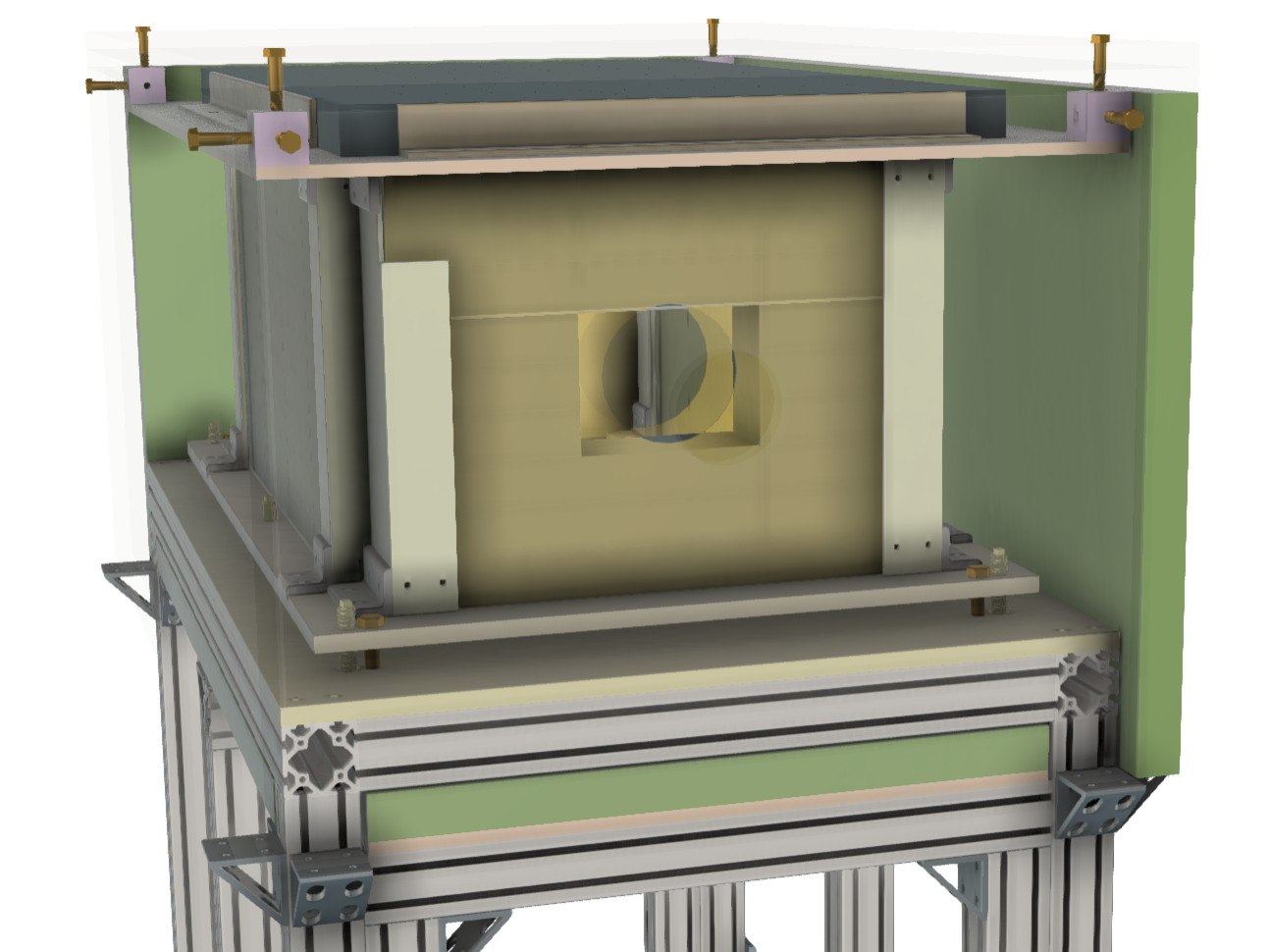}
\caption{Left: Outermost view of the initial designs for the NaI stand. Right: Initial design of the NaI stand inside of the two layers of borated polyethylene and lead.}
\label{fig:Detector design photo}
\end{figure*}

The stand and radiation shielding for the detectors pictured in Fig. ~\ref{fig:Detector design photo} was designed at IU and assembled at LANSCE. It consisted mostly of 80/20 support rails, borated polyethylene, lead bricks, and sheets of aluminum. The borated polyethylene and lead bricks were present to prevent both neutrons scattered in the cave and gammas from neutron capture in the cave walls from making it to the detectors.

A 48.125" long polarized neutron spin-transport tube with outer diameter 4" was placed in the center of the array, holding the targets. It consisted of an aluminum tube wrapped in 10 AWG copper wire, producing a magnetic field in the $z$ direction throughout the detector array to match the direction and strength at the entry of the array produced from the neutron spin flipper. Stray magnetic fields from the spin-transport tube were measured to have a negligible effect on the PMTs. The spin-transport tube was wrapped with polytetrafluoroethylene film tape to prevent possible electrical shorts. 

\begin{figure}[!h]
\centering
\includegraphics[width=0.48\textwidth]{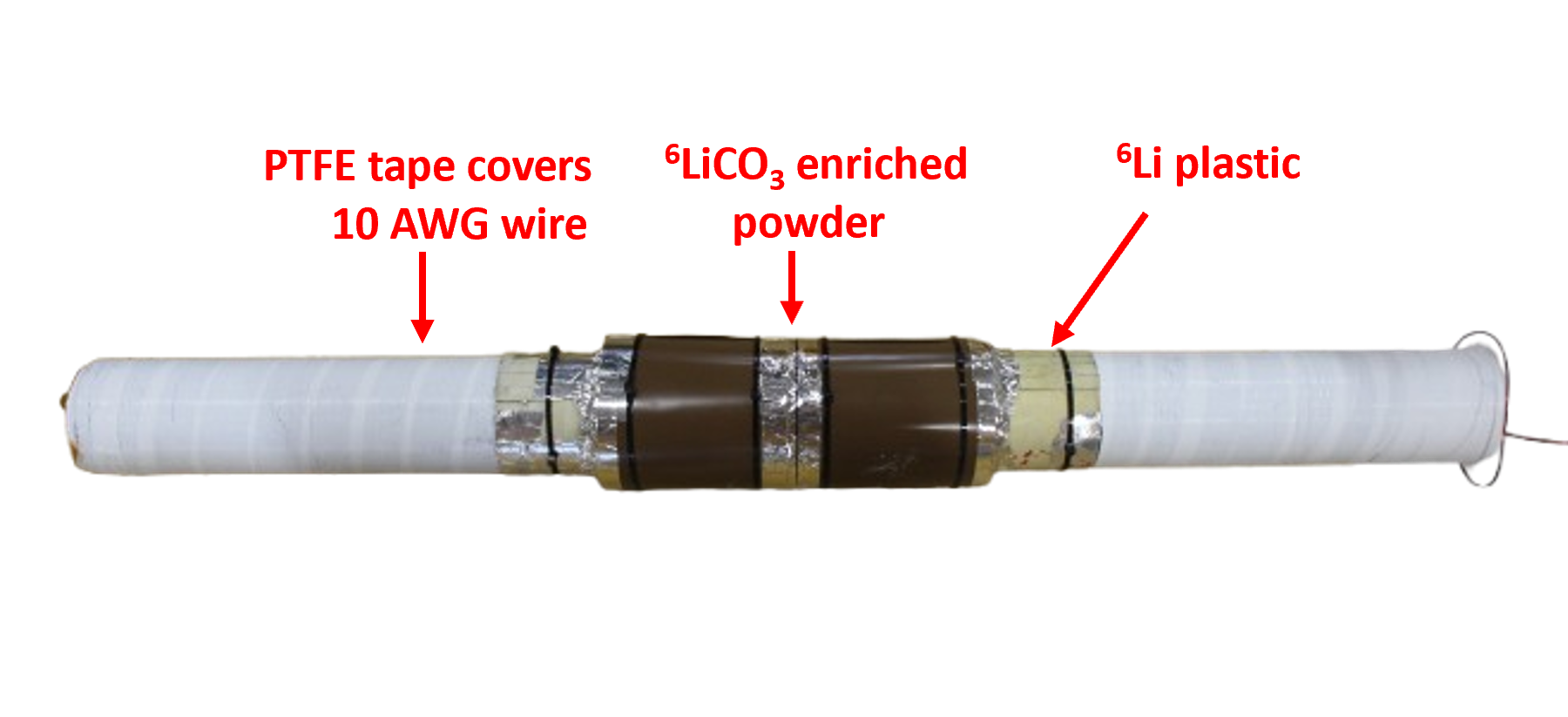}
\caption{Spin-transport tube before insertion into the NaI array.}
\label{fig:spin transport tube}
\end{figure}

To prevent neutrons from scattering off of the target and into the gamma array (and thereby slowly activating the detectors from neutron capture in Na and I), we surrounded the target region with $^{6}$Li-rich material, which creates the smallest fraction of gamma rays per incident neutron capture of any nucleus of an element easily used in solid form. The center of the array was wrapped with $^6$Li-enriched lithium plastic, which was 49.8 cm in length. Wrapped around the lithium plastic was 1,300 g of 95.56\% $^6$LiCO$_3$ double wrapped in anti-static bags 13" in length. Finally, the bags were wrapped with aluminum strips and taped in place with aluminum foil to prevent them from moving in $z$. The final assembly of the spin-transport tube before insertion into the array is shown in Fig.~\ref{fig:spin transport tube}.

\subsection{P-odd Asymmetry Observation in $^{139}$La}

\begin{figure}[!]
    \includegraphics[width=1.0\linewidth]{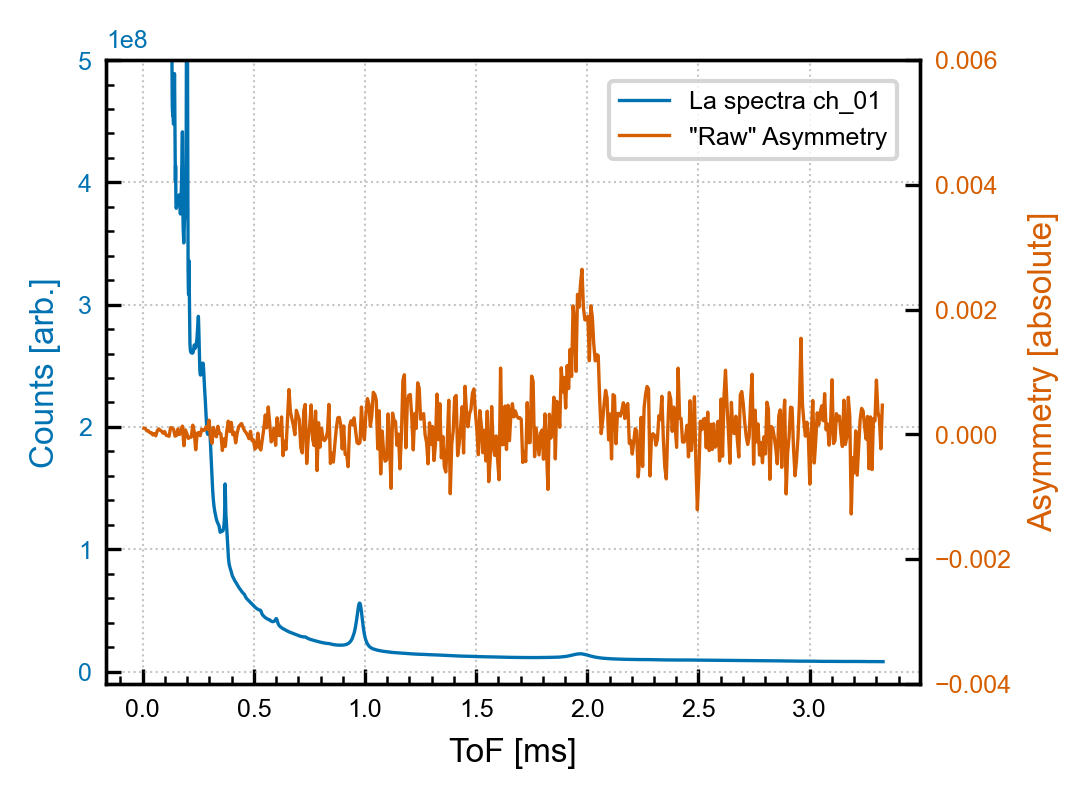}
    \caption{A view of the La spectrum (left axis), with the ``raw" asymmetry superimposed (right axis). There is a clear asymmetry over the p-wave resonance at 1.96 ms, and none at the s-wave resonance at 0.98 ms, as expected.}
    \label{fig:prelim_asym}
\end{figure}

Approximately 75 hours of current-mode data with polarized beam was taken with the NaI detector array on a La target. Data was amassed and separated into up and down neutron spin states to measure the PV asymmetry. A simple analysis with no background or resonance fitting can be done by measuring the asymmetry in the spectra of these two spin states to confirm that the apparatus can see a known parity-odd asymmetry and that there are no obvious systematic errors on resonances with no P-odd effects. Fig. \ref{fig:prelim_asym} shows a clear P-odd asymmetry at the 0.7 eV p-wave resonance in the La target, and therefore the ability of the NaI detector array to detect parity violation. The details of the full experimental apparatus and results of this PV search are the work of a future paper.

\section{Conclusion and Future Steps}

Based on the results of the JPARC test run and LANSCE beamtime, this current-mode NaI(Tl) gamma detector array is capable of resolving p-wave resonances in heavy nuclei, and can be used in future PV resonance searches. The NOPTREX collaboration plans to use this array to search for new PV resonances in various heavy nuclei at LANSCE. The high level of modularity in the design of the array will also allow for it to be used in other experimental efforts in search of CP symmetry violation. 

This array can be used for additional physics beyond the parity violation searches discussed above. Neutron-nucleus resonance cross sections in the isolated resonance regime of interest for our work fall into the kinematic regime $kR<<1$, where $k$ is the neutron wave vector and $R$ is the range of the neutron-nucleus strong interaction. The amplitude for neutron-nucleus resonances for different values of the orbital angular momentum $L$ in conventional scattering theory, in which the wave packet of the incident neutron is simply a weighted superposition of plane waves of slightly different $\vec{k}$ over a narrow range $\delta \vec{k}$, is suppressed by a factor of $(kR)^{L}$. If however it were possible to create a ``twisting" neutron wave packet with nonzero angular momentum, one could overcome the traditional angular momentum barrier in neutron-nucleus scattering and perhaps increase the resonance amplitudes of nonzero $L$. Ongoing research into the production of neutron beams with orbital angular momentum wave packets may succeed to create such beams in the future. Our array is well-suited to search for small changes in p-wave neutron-nucleus resonance cross sections.    

\section{Acknowledgments}
\label{sec:ack}

J. T. Mills, J. G. Otero Munoz, K. Dickerson, J. Doskow, M. Luxnat, S. Samiei, W. M. Snow, D. Schaper, and G. Visser acknowledge support from US National Science Foundation grant PHY-2209481, from the U.S. Department of Energy, Office of Science,  DOE Quantum Horizons Initiative award number DE‐SC0023695, and from the Indiana University Center for Spacetime Symmetries. J. Fry, J. T. Mills, A. Quintinar-Pen\~na, A. Richburg, and D. Slone acknowledge support from US National Science Foundation grant 1849213, sub-award 3200002692. K. Dickerson acknowledges support from  the DOE SCGSR program and from the GNeUS Global Neutron Scientists Fellowship of the European Union  Horizon 2020 Research and Innovation Programme. J. G. Otero Munoz acknowledges support from the NSF AGEP program, the GEM fellowship program, and the DOE SCGSR program. The beamtime for the test measurement at JPARC was awarded through the JPARC Long-Term Proposal Award 2020L1300. We would like to acknowledge the scientists at the Rexon Components, Inc. (John Sheehan, Zaid Farukhi, and M. R. Farukhi) for their successful revival of the old NaI(Tl) crystals we used for this array and for their design and fabrication of the detector encapsulation and scintillation light transport. This research used resources provided by the Los Alamos National Laboratory Los Alamos Neutron Science Center (LANSCE), which is supported by the U.S. Department of Energy National Nuclear Security Administration under Contract No. 89233218CNA000001.

\clearpage
\bibliographystyle{elsarticle-num}
\bibliography{ref.bib}

@techreport{kulesza2022mcnp,
  title={MCNP{\textregistered} code version 6.3. 0 theory \& user manual},
  author={Kulesza, Joel A and Adams, Terry R and Armstrong, Jerawan Chudoung and Bolding, Simon R and Brown, Forrest B and Bull, Jeffrey S and Burke, Timothy Patrick and Clark, Alexander Rich and Forster III, Robert Arthur Art and Giron, Jesse Frank and others},
  year={2022},
  institution={Los Alamos National Laboratory (LANL), Los Alamos, NM (United States)}
}

@article{seestrom1999apparatus,
  title={Apparatus for parity-violation study via capture $\gamma$-ray measurements},
  author={Seestrom, SJ and Frankle, CM and Bowman, JD and Crawford, BC and Haseyama, T and Masaike, A and Matsuda, A and Penttil{\"a}, SI and Roberson, RN and Sharapov, EI and others},
  journal={Nuclear Instruments and Methods in Physics Research Section A: Accelerators, Spectrometers, Detectors and Associated Equipment},
  volume={433},
  number={3},
  pages={603--613},
  year={1999},
  publisher={Elsevier}
}

@article{masuda1989longitudinal,
  title={Longitudinal asymmetry in a neutron radiative capture reaction of 139La},
  author={Masuda, Y and Adachi, T and Masaike, A and Morimoto, K},
  journal={Nuclear Physics A},
  volume={504},
  number={2},
  pages={269--276},
  year={1989},
  publisher={Elsevier}
}

@article{Gericke:2004xn,
    author = "Gericke, M. T. and others",
    title = "{A Current mode detector array for gamma-ray asymmetry measurements}",
    eprint = "nucl-ex/0411022",
    archivePrefix = "arXiv",
    reportNumber = "LA-UR-1055",
    doi = "10.1016/j.nima.2004.11.043",
    journal = "Nucl. Instrum. Meth. A",
    volume = "540",
    pages = "328--347",
    year = "2005"
}

@article{YEN2000476,
title = {A high-rate 10B-loaded liquid scintillation detector for parity-violation studies in neutron resonances},
journal = {Nuclear Instruments and Methods in Physics Research Section A: Accelerators, Spectrometers, Detectors and Associated Equipment},
volume = {447},
number = {3},
pages = {476-489},
year = {2000},
issn = {0168-9002},
doi = {https://doi.org/10.1016/S0168-9002(99)01308-X},
url = {https://www.sciencedirect.com/science/article/pii/S016890029901308X},
author = {Yi-Fen Yen and J.D. Bowman and R.D. Bolton and B.E. Crawford and P.P.J. Delheij and G.W. Hart and T. Haseyama and C.M. Frankle and M. Iinuma and J.N. Knudson and A. Masaike and Y. Masuda and Y. Matsuda and G.E. Mitchell and S.I. Penttilla and N.R. Roberson and S.J. Seestrom and E. Sharapov and H.M. Shimizu and D.A. Smith and S.L. Stephenson and J.J. Szymanski and S.H. Yoo and V.W. Yuan}
}

@article{SZYMANSKI1994564,
title = {Ion chamber system for neutron flux measurements},
journal = {Nuclear Instruments and Methods in Physics Research Section A: Accelerators, Spectrometers, Detectors and Associated Equipment},
volume = {340},
number = {3},
pages = {564-571},
year = {1994},
issn = {0168-9002},
doi = {https://doi.org/10.1016/0168-9002(94)90139-2},
url = {https://www.sciencedirect.com/science/article/pii/0168900294901392},
author = {J.J. Szymanski and J.D. Bowman and P.P.J. Delheij and C.M. Frankle and J. Knudson and S. Penttilla and S.J. Seestrom and S.H. Yoo and V.W. Yuan and X. Zhu}
}

@article{SEESTROM1999603,
title = {Apparatus for parity-violation study via capture gamma ray measurements},
journal = {Nuclear Instruments and Methods in Physics Research Section A: Accelerators, Spectrometers, Detectors and Associated Equipment},
volume = {433},
number = {3},
pages = {603-613},
year = {1999},
issn = {0168-9002},
doi = {https://doi.org/10.1016/S0168-9002(99)00481-7},
url = {https://www.sciencedirect.com/science/article/pii/S0168900299004817},
author = {S.J. Seestrom and C.M. Frankle and J.D. Bowman and B.C. Crawford and T. Haseyama and A. Masaike and A. Matsuda and S.I. Penttilla and R.N. Roberson and E.I. Sharapov and S.L. Stephenson}
}

@article{n3He:2020zwd,
    author = "Gericke, M. T. and others",
    collaboration = "n3He",
    title = "{First Precision Measurement of the Parity Violating Asymmetry in Cold Neutron Capture on $^3$He}",
    eprint = "2004.11535",
    archivePrefix = "arXiv",
    primaryClass = "nucl-ex",
    doi = "10.1103/PhysRevLett.125.131803",
    journal = "Phys. Rev. Lett.",
    volume = "125",
    number = "13",
    pages = "131803",
    year = "2020"
}

@article{PENN2001332,
title = {A low-noise 3He ionization chamber for measuring the energy spectrum of a cold neutron beam},
journal = {Nuclear Instruments and Methods in Physics Research Section A: Accelerators, Spectrometers, Detectors and Associated Equipment},
volume = {457},
number = {1},
pages = {332-337},
year = {2001},
issn = {0168-9002},
doi = {https://doi.org/10.1016/S0168-9002(00)00748-8},
url = {https://www.sciencedirect.com/science/article/pii/S0168900200007488},
author = {S.D. Penn and E.G. Adelberger and B.R. Heckel and D.M. Markoff and H.E. Swanson}
}

@article{Vesna:2021rma,
    author = "Vesna, V. A. and Gledenov, Yu. M. and Nesvizhevsky, V. V. and Sedyshev, P. V. and Shulgina, E. V.",
    title = "{Search and Measurement of $P$-Odd Asymmetry in the Emission of Secondary Particles Produced in Reactions of Polarized Thermal and Cold Neutrons with Light Nuclei}",
    doi = "10.1134/S1063779621010044",
    journal = "Phys. Part. Nucl.",
    volume = "52",
    number = "1",
    pages = "1--18",
    year = "2021"
}

@article{BOWMAN1990183,
title = {Current-mode detector for neutron time-of-flight studies},
journal = {Nuclear Instruments and Methods in Physics Research Section A: Accelerators, Spectrometers, Detectors and Associated Equipment},
volume = {297},
number = {1},
pages = {183-189},
year = {1990},
issn = {0168-9002},
doi = {https://doi.org/10.1016/0168-9002(90)91365-I},
url = {https://www.sciencedirect.com/science/article/pii/016890029091365I},
author = {J.D. Bowman and J.J. Szymanski and V.W. Yuan and C.D. Bowman and A. Silverman and X. Zhu}
}

@article{sharapov1991capture,
  title={Capture Gamma-Ray Spectroscopy, edited by RW Hoff AIP},
  author={Sharapov, EI and Wender, SA and Postma, H and Seestrom, SJ and Gould, CR and Wasson, OA and Popov, Yu P and Bowman, CD},
  journal={New York},
  pages={756},
  year={1991}
}

@article{Schaper2020,
  title={A modular apparatus for use in high-precision measurements of parity violation in polarized eV neutron transmission},
  author={Schaper, Danielle C and Auton, Clayton and Barr{\'o}n-Palos, Libertad and Borrego, M and Chavez, Andrew and Cole, Lillie and Crawford, Christopher B and Curole, Jonathan and Dhahri, Hajer and Dickerson, Kelly A and others},
  journal={Nuclear Instruments and Methods in Physics Research Section A: Accelerators, Spectrometers, Detectors and Associated Equipment},
  volume={969},
  pages={163961},
  year={2020},
  publisher={Elsevier},
  doi = {https://doi.org/10.1016/j.nima.2020.163961}
}

@article{potter1974test,
  title={Test of Parity Conservation in p- p Scattering},
  author={Potter, JM and Bowman, JD and Hwang, CF and McKibben, JL and Mischke, RE and Nagle, DE and Debrunner, PG and Frauenfelder, H and Sorensen, LB},
  journal={Physical Review Letters},
  volume={33},
  number={21},
  pages={1307},
  year={1974},
  publisher={APS}
}

@article{yuan1986measurement,
  title={Measurement of parity nonconservation in the proton-proton total cross section at 800 MeV},
  author={Yuan, V and Frauenfelder, H and Harper, RW and Bowman, JD and Carlini, R and MacArthur, DW and Mischke, RE and Nagle, DE and Talaga, RL and McDonald, AB},
  journal={Physical Review Letters},
  volume={57},
  number={14},
  pages={1680},
  year={1986},
  publisher={APS}
}

@article{adelberger1985parity,
  title={Parity violation in the nucleon-nucleon interaction},
  author={Adelberger, EG and Haxton, WC},
  journal={Annual review of nuclear and particle science. Volume 35},
  pages={501--558},
  year={1985}
}

@article{alfimenkov1983parity,
  title={Parity nonconservation in neutron resonances},
  author={Alfimenkov, VP and Borzakov, SB and Van Thuan, Vo and Mareev, Yu D and Pikelner, LB and Khrykin, AS and Sharapov, EI},
  journal={Nuclear Physics A},
  volume={398},
  number={1},
  pages={93--106},
  year={1983},
  publisher={Elsevier}
}

@article{bunakov1981parity,
  title={Parity non-conservation effects in neutron elastic scattering reactions},
  author={Bunakov, VE and Gudkov, VP},
  journal={Zeitschrift f{\"u}r Physik A Atoms and Nuclei},
  volume={303},
  number={4},
  pages={285--291},
  year={1981},
  publisher={Springer}
}

@article{bunakov1983parity,
  title={Parity violation and related effects in neutron-induced reactions},
  author={Bunakov, VE and Gudkov, VP},
  journal={Nuclear Physics A},
  volume={401},
  number={1},
  pages={93--116},
  year={1983},
  publisher={Elsevier}
}

@misc{PMT_base_circuitdiagram,
  author       = {Visser, Gerard and
                  Otero Munoz, Jorge G. and
                  Mills, J.T. and
                  Snow, W.M.},
  title        = {NOPTREX Custom PMT Base Circuit Diagram},
  month        = apr,
  year         = 2026,
  publisher    = {Zenodo},
  doi          = {10.5281/zenodo.19361052},
  url          = {https://doi.org/10.5281/zenodo.19361052},
}

\end{document}